\documentstyle[12pt,fleqn,aasms4]{article}

\newcommand{\be}{\begin{equation}}
\newcommand{\ee}{\end{equation}}
\newcommand{\bea}{\begin{eqnarray}}
\newcommand{\eea}{\end{eqnarray}}

\newcommand{\kms}{{\rm km}\,{\rm s}^{-1}}
\def\lmc{{\rm LMC}}

\begin{document}

\title{Systematics of RR Lyrae Statistical Parallax III: 
Apparent Magnitudes and Extinctions}
\author
{Andrew Gould\altaffilmark{1}\altaffiltext{1}{Alfred P.\ Sloan Foundation Fellow} and Piotr Popowski\altaffilmark{2}\altaffiltext{2}{Ohio State University Presidential
Fellow}}
\affil{Ohio State University, Department of Astronomy, Columbus, OH 43210}
\affil{E-mail: gould, popowski@astronomy.ohio-state.edu}

\begin{abstract}

We sing the praises of the central limit theorem.  Having previously 
removed all other possible causes of significant systematic error
in the statistical parallax determination of RR Lyrae absolute magnitudes,
we investigate systematic errors from two final sources of input data: 
apparent magnitudes and extinctions.  We find corrections due to each of
about 0.05 mag, i.e., about half the statistical error.  However, these
are of opposite sign and so approximately cancel out.  The apparent magnitude
system that we previously adopted from Layden et al.\ was calibrated to
the photoelectric photometry of Clube \& Dawe.  Using Hipparcos photometry and
archival modern ground-based photometry, we show that the Clube \& Dawe system
is about 0.05 mag too bright.  Extinctions were previously based on the map
of Burstein \& Heiles which was constructed from HI maps.  We argue that 
extinctions should rather be estimated using the new map of 
Schlegel, Finkbeiner \& Davis based on {\it COBE} and {\it IRAS} measurements
of dust emission.  This substitution increases the mean estimated extinction
by about 0.05 mag, primarily because of a difference in the zero point of the 
two maps.  Our final estimate for the absolute magnitude is
$M_V=0.77\pm0.13$ at [Fe/H]$=-1.60$ for a pure sample of 147 halo 
RR Lyrae stars, or $M_V=0.80\pm 0.11$ 
at [Fe/H]$=-1.71$ if we incorporate kinematic information
from 716 non-kinematically selected non-RR Lyrae stars from Beers \& 
Sommer-Larsen.  These are $2\sigma$ and $3\,\sigma$ fainter than recent 
determinations
of $M_V$ based on main-sequence fitting of clusters using Hipparcos 
measurements of subdwarfs by Reid and Gratton et al.  Since statistical 
parallax is being cleared of systematic errors and since the probability of
a more than $2\,\sigma$ statistical fluctuation is less than 1/20, 
we conclude that
these brighter determinations may be in error.  In the course
of these three papers, we have corrected 6 systematic errors whose 
absolute values total 0.20 mag.  Had these, contrary to the expectation
of the central limit theorem, all lined up one way, they could have resolved
the conflict in favor of the brighter determinations.  In fact, the net 
change was only 0.06 mag.

\keywords{astrometry --- distance scale --- extinction--- 
Galaxy: kinematics and dynamics 
--- methods: analytical, statistical --- stars: variables: RR Lyrae}
\end{abstract}

\section{Introduction}

	Statistical parallax appears to be an extremely robust method for
measuring the absolute magnitude of halo RR Lyrae stars.  Nevertheless, the
results of this method are in serious conflict with several other 
determinations.  In Paper II of this series (Popowski \& Gould 1998b)
we found
\be
M_V=0.74\pm 0.12,\quad {\rm at}\ \left<\rm [Fe/H]\right>=-1.60 \qquad 
({\rm pure}\ {\rm RR}\ {\rm Lyrae}),\label{eqn:pureRR}
\ee
for a sample of 165 halo RR Lyrae stars with high-quality proper motions
from the Hipparcos (ESA 1997) and Lick NPM1 
(Klemola, Hanson, \& Jones 1993) surveys.  We also combined this
result with a separate determination based on a non-kinematically selected
sample of 103 RR Lyrae stars and 724 non-RR Lyrae stars from 
Beers \& Sommer-Larsen (1995) and (taking account of the 0.45 correlation
coefficient between the two samples) found
\be
M_V=0.77\pm 0.10,\quad {\rm at}\ \left<\rm [Fe/H]\right>=-1.71 \qquad 
({\rm combined}).\label{eqn:combined}
\ee
The former value can be compared with measurements based on main-sequence
fitting of globular clusters to subdwarfs with Hipparcos parallaxes
which yields  $M_V\sim 0.44\pm 0.08$ (Reid 1997) or $M_V\sim 0.49 \pm 0.04$ 
(Gratton et al.\ 1997; Gratton 1998) at the same metallicity.  
(These comparisons take
account of differences in the metallicity scales used by different authors
as we discuss more fully in the Appendix.)\ \  If equation 
(\ref{eqn:pureRR}) is combined with the measurement of the dereddened 
apparent magnitude of RR Lyrae stars in the Large Magellanic Cloud (LMC)
of $V_0 = 18.98\pm 0.05$ (Hazen \& Nemec 1992; Popowski \& Gould 1998a
-- Paper I), this yields a distance modulus $\mu_\lmc = 18.24\pm 0.14$.
(Here we have assumed an LMC metallicity [Fe/H]$=-1.8$, and a slope
$M_V =$const $+0.15$[Fe/H], but the exact value of the slope makes very 
little difference because the metallicities in eq.\ \ref{eqn:pureRR} and
of the LMC are so similar.)\ \  This result is quite low compared to the
``traditional'' value $\mu_\lmc = 18.50$ and is even lower compared to those
derived using Hipparcos-based calibrations of RR Lyrae stars and Cepheids:
$\mu_\lmc = 18.65\pm 0.1$ (Reid 1997), $\mu_\lmc = 18.63\pm 0.06$ 
(Gratton et al.\ 1997), and $\mu_\lmc = 18.70\pm 0.10$ (Feast \& Catchpole 
1997).

	In principle, these discrepancies could be due to a 
greater than $2\,\sigma$
statistical fluctuation.  However, for Gaussian statistics, the probability
of a $2\,\sigma$ fluctuation is $<1/20$. (Moreover, for the statistical 
parallax determination, we have checked that the distribution of errors has
Gaussian tails, even when the input data are not Gaussian distributed.)\ \
The usual cause of $>2\,\sigma$ discrepancies is not statistical
fluctuations but systematic errors, and one is therefore led to suspect that
there are unrecognized systematic errors in one or several of these 
measurements.  Moreover, the conflict with equation (\ref{eqn:combined})
is even stronger, about $3\,\sigma$.  While there are some additional 
assumptions that go into equation (\ref{eqn:combined}) that make it
overall less robust than equation (\ref{eqn:pureRR}), the combined 
determination nevertheless argues against a large statistical fluctuation
as the source of the discrepancy.

	This is the third and final paper in a series designed to essentially
eliminate the possibility of a significant systematic error in the
statistical parallax determination.  Statistical parallax works in effect
by forcing equality between the velocity ellipsoids as determined from
radial velocities, and from proper motions.  That is, one can measure the nine
parameters describing the velocity ellipsoid (three components of bulk motion,
$w_i$, plus
six independent components of the velocity-dispersion tensor, $C_{i j}$) 
from radial
velocities alone.  On the other hand, if 
one {\it assumes} some arbitrary absolute magnitude
for the RR Lyrae stars, then one can infer their distances from their
measured apparent magnitudes and estimated extinctions.  The distances and
proper motions yield the transverse velocities, and from these one can
again estimate the nine parameters of the velocity ellipsoid.  One could
then adjust the assumed absolute magnitude so that the velocity ellipsoid
from proper motions matched the velocity ellipsoid from radial velocities
as closely as possible.  In practice, one fits for all 10 parameters
(nine for the velocities plus the absolute magnitude) simultaneously
using maximum likelihood.

	Logically, there are three possible ways for systematic errors
to enter the determination.  First, the {\it mathematics} of the method 
itself could introduce biases.  Second, the RR Lyrae sample could fail to 
satisfy some of the {\it physical} properties assumed by the method.
Third, one or more of the four major {\it observational inputs} 
(proper motions, 
radial velocities, apparent magnitudes, and extinctions) could be 
systematically in error.  (A fifth observational input, metallicities,
requires a separate discussion.  Different studies may be on 
systematically different metallicity scales, and care must therefore
be exercised when comparing the results from these investigations.  
See Appendix.)\ \ 

	In Paper I, we investigated possible systematic errors arising
from the mathematical method and physical assumptions.  An example of a
potential mathematical problem is that the
likelihood method explicitly assumes that the velocity distribution is 
Gaussian while, as we showed, the actual distribution is highly non-Gaussian.
An example of a potential physical problem is that the method implicitly
assumes that the velocity-dispersion tensor does not depend on location
despite the fact that the stars are found at distances ($\la 2\,$kpc) that
are a significant fraction of the Galactocentric distance ($R_0\sim 8\,$kpc).
We examined a large number of such effects, some by vigorous Monte Carlo
simulations and some with the aid of mathematical arguments.  We corrected
for all of them although most were smaller than 0.01 mag and tended to 
mutually cancel one another.  The largest correction (0.03 mag fainter) was 
due to Malmquist bias
which had been previously recognized but not previously incorporated into the
analysis.

	In Paper II, we investigated systematic errors arising from the
first two observational inputs, proper motions and radial velocities.
The proper motions are of greater concern because they are intrinsically
more difficult to measure and hence have larger fractional errors.  We had
already noted in Paper I that if the proper-motion {\it errors} are 
misestimated, this can introduce significant systematic errors even if the 
proper motion themselves are unbiased.  We used the precise Hipparcos 
proper motions (when available) to test the two large catalogs, Lick and
Wan, Mao, \& Ji (1980, WMJ), that had previously been used and found that
indeed the Lick errors had been slightly underestimated and the WMJ errors
had been seriously underestimated.  These two corrections moved $M_V$
brighter by 0.04 mag, but this was mostly compensated by random changes
induced by substituting the more precise Hipparcos proper motions (when
available) for the previous values.  We also tested all three catalogs
to search for non-statistical errors and removed five questionable stars.

	Radial velocities are in principle much easier to measure than
proper motions.  However, for pulsating variables, the measured velocity of
(the atmosphere of) the star can differ from its center of mass by 
$\sim 50\,\kms$ and hence an accurate velocity determination requires many
measurements and/or good phasing.  The quality of the radial velocity data
varies from star to star, and it was therefore possible that the errors had
been either systematically overestimated or underestimated.  In Paper II, we
checked the entire system of the radial-velocity measurements
by, in effect, determining the
radial-velocity ellipsoid from the Beers \& Sommer-Larsen (1995)
 non-kinematically selected sample of metal-poor {\it non-RR Lyrae} halo
stars.  The resulting
$M_V$ was consistent with the one derived from the pure RR Lyrae sample, 
indicating that the radial velocities are not a source of significant
systematic error.  

	In brief, Paper I checked for and removed all sources of systematic 
error coming from the mathematics of the method and the physical assumptions
about the sample, down to a level well below the statistical error.  Paper II
did the same for two of the observational inputs: proper motions and radial
velocities.

	Here we turn our attention to the remaining two observational inputs:
apparent magnitudes and extinctions.  At first sight, it does not seem that
there could be much controversy about the apparent magnitude of $V\sim 12$
stars.  However, exactly because the stars are bright, many were measured
long ago.  Layden et al.\ (1996) compiled photometric measurements 
from several sources
and attempted to put them on a common system aligned with their large subsample
from Clube \& Dawe (1980) which has photoelectric photometry and which they
assumed to be equivalent to the modern (Landolt 1992) system.  In particular,
they found the photoelectric photometry of Bookmeyer et al.\ (1977) to be
on average 0.06 mag fainter than that of Clube \& Dawe (1980) and transformed 
it accordingly (see Layden et al.\ 1996, Table 1).  Thus, there are
uncertainties in the apparent-magnitude scale of order 0.06 mag which, 
according to equation (\ref{eqn:pureRR}), is half the size of
the statistical error.  In \S\ 3, we test the Layden et al.\ (1996) system 
against Hipparcos photometry.  We show that for $V\la 12$, the 
{\it untransformed} Bookmeyer et al.\ (1977) photometry is in good 
agreement with Hipparcos.  The Clube \& Dawe (1980) photoelectric
photometry also agrees
well with Hipparcos for $V\la 10.5$ 
but is systematically brighter than Hipparcos
by $\sim 0.06$ 
mag for $11\la V \la 12$.  The most straight forward interpretation
of these results is that the Bookmeyer et al.\ (1977) photometry
is more reliable than the Clube \& Dawe (1980) photometry and that 
therefore the Layden et al.\ (1996) system is too bright by about 0.06 mag.
This conclusion is confirmed by the good agreement between Hipparcos and the
high quality photometry of Jones et al.\ (1992), Schmidt (1991), and 
Schmidt, Chab, \& Reiswig (1995).

	Extinctions pose another set of problems.
For stars that are far from the Galactic plane, one can assume that they 
are above essentially all of the dust along their line of sight.  
One can therefore adopt the extinctions as measured for extragalactic objects
along the same (or very nearby) lines of sight.  Burstein \& Heiles (1982, BH)
have constructed a map of such extinctions over a large fraction of the sky
by combining galaxy counts and HI measurements.  Layden et al.'s (1996) 
extinction
estimates are based primarily on this map for the great majority of the sample.
However, there are some lines of sight (particularly at low latitudes)
for which BH do not give extinctions and others
where the star is relatively close to the plane so that some of the dust
may lie behind the star.  In the latter cases, the BH
map would overestimate the extinctions.  For these stars, Layden et al.\ (1996)
adopted other methods to estimate the extinction, notably the colors of the
stars.  Since the intrinsic color of RR Lyrae stars is a function of the 
period with relatively little scatter, this method should work well at least
on average.  However, Sturch (1966) had earlier used colors
to estimate the extinctions toward a large sample of RR Lyrae stars, and Layden
et al.\ (1996) 
found that these estimates were systematically higher than their BH-based
values by 0.11 mag.  Layden et al.\
(1996) argued that the BH-based system was correct,
and attempted to put stars without BH extinctions on the same BH system.
Nevertheless, it is important to note that if the Sturch (1966) system were
correct, the dereddened apparent magnitudes would be systematically brighter
by 0.11 mag, and hence the absolute-magnitude estimate would be brighter
by the same amount.  This would move the statistical-parallax estimates of
$M_V$ and $\mu_\lmc$ closer by $1\,\sigma$ to the estimates obtained using 
competing methods, and would thus help significantly to resolve the 
controversy.

	In \S\ 4, we therefore re-evaluate the extinctions using
a different approach.  First, we base our determinations on the
new extinction map of Schlegel, Finkbeiner, \& Davis (1998, SFD).  
We argue that the SFD map is superior to the BH map both in its level of 
detail and in its zero point.  (The zero point of SFD is about 0.06 mag higher
in $A_V$ than the BH map.)\ \ Second, we restrict attention to 
stars that
are more than 300 pc from the Galactic plane.  These lie beyond most of the
dust and therefore the SFD extinctions require only small corrections.  
Third, we exclude the four stars with SFD extinctions
$A_V>0.56$
since comparison with Layden et al.\ (1996) shows a systematic deviation
for these stars and we are unable to determine which system is in error.
We find a correction due to revised extinctions which makes $M_V$ about
0.05 mag brighter.  Combining the corrections
due to revised apparent magnitudes and extinctions, we find that
equations (\ref{eqn:pureRR}) and (\ref{eqn:combined}) are each increased
(made fainter) by 0.03 mag.  These changes are in fact mainly due to
random fluctuations caused by the fact that we are using a slightly
different sample (147 vs.\ 165 stars).  The two systematic effects that
we identify here almost precisely cancel out.
Our results are all presented in \S\ 5,
and we discuss the implications of
these results in \S\ 6.  We begin in \S\ 2 by describing our basic sample.

\section{Sample}

	As in Paper II, our initial sample of type ab RR Lyrae stars 
comes from two sources.  First, Layden et al.\ (1996) give proper motions,
radial velocities, apparent magnitudes, extinctions, and metallicities for
213 stars.  Second, Hipparcos gives proper motions for many of these 213
and for 19 additional stars for which there are radial velocities, 
apparent magnitudes, extinctions, and metallicities in Layden (1994).  
These are
VX Scl, SX For, RX Col, HH Pup, RV Oct, TY Aps, XZ Aps, RW Tra, 
WY Pav, MS Ara, IN Ara, V455 Oph, V413 CrA, BK Dra, BN Pav, BP Pav, Z Mic, 
RY Oct, and SS Oct.  In Paper II, we also included in our master file 
another star, BX Dra, which is probably an eclipsing variable, not an RR 
Lyrae star (Fernley et al.\ 1998).  However, this was classified as 
a disk star and so did not enter the final sample.
In Paper II, we eliminated a number of specific
stars because of doubts about the quality of their proper motions.  These
included all stars with only WMJ proper motions.  That is, our sample was
composed entirely of stars with Lick and/or Hipparcos proper motions.  We
repeat this procedure in the current paper.

	As in Paper II, we consider two different samples.  First, 
we obtain a pure RR Lyrae sample of stars belonging to ``halo-3''
as defined by Layden et al.\ (1996).  Second, we select a non-kinematic
sample of {\it both} RR Lyrae stars and non-RR Lyrae stars using a metallicity
criterion, [Fe/H]$<-1.5$.
The latter are drawn from the sample of 1836 stars of Beers \& Sommer-Larsen
(1995).  As we discuss in \S\ 3 and \S\ 4, we slightly modify the selection 
procedures of Paper II to account for new apparent-magnitude and extinction
information.

\section{Apparent Magnitudes}

	Equations (\ref{eqn:pureRR}) and (\ref{eqn:combined}) were derived
making use of photometry compiled by Layden et al.\ (1996, L96) when available,
and in a few cases from Layden (1994, L94) 
(which ultimately has the same sources).
Of the 213 stars (including 162 ``halo-3'' stars) in L96, 
57 (51 ``halo-3'') have photoelectric
photometry from Clube \& Dawe (1980, CD80), 81 (54) have photoelectric
photometry from
Bookmeyer et al.\ (1977, B77), 7 (4) have Walraven photometry from Lub (1977), 
21 (14) have CCD photometry from Schmidt (1991, S91) and Schmidt et al.\ 
(1995, S95), 8 (5) have preliminary photoelectric 
photometry from L96, and 39 (34) have
photometry compiled by L94 from heterogeneous sources in the
General Catalog of Variable Stars (Kholopov 1985, GCVS).  L96
attempted to place all of these on a common system aligned
with the photoelectric photometry of CD80.

	In particular, they transformed the B77 photoelectric photometry and
the S91 and S95 CCD photometry according to
\be
V_{\rm L96} = 0.146 + 0.983 V_{\rm B77}
\label{eqn:bookmeyer}
\ee
and
\be
V_{\rm L96} = -0.086 + V_{\rm S91,S95}.
\label{eqn:schmidt}
\ee

	Fernley et al.\ (1998) have fit the light curves of 144 RR Lyrae
stars (including 123 type ab and 21 type c) using photometry data from 
Hipparcos.  They transformed the Hipparcos $V_H$ magnitudes into Johnson $V$
magnitudes using $V = V_H - X$ where $X=0.09$ for type ab and 
$X=0.06$ for type c, in accordance with the transformations given
in the Hipparcos catalog at the appropriate colors for type ab and type c,
respectively.  They tested these values against the precise ground-based
measurements of Liu \& Janes (1990) for 13 stars with intensity weighted
means in the range $9.5\la V \la 11.1$
including 11 type ab and 2 type c.  They found a mean offset of only 0.003 mag,
with a scatter of 0.007 mag.  This gives high confidence both in the 
underlying Hipparcos data and in Fernley et al.'s (1998, F98) procedure for 
recovering 
intensity-weighted means.  We therefore propose to use the F98
data set to
test the photometry of the larger L96 compilation.

	Before doing so, however, we first note that the Hipparcos 
magnitudes require small additional corrections due to extinction.
According to the Hipparcos catalog, in the neighborhood of 
$V - V_H\sim -0.09$, $d(V -V_H)/d(V-I) =-0.16$ and therefore,
for reddened RR Lyrae stars, one must further adjust the Hipparcos magnitudes
by $-0.16 E(V-I)\sim -0.2 E(B-V)$.  This results in our final estimate for 
Hipparcos-based Johnson $V$ magnitudes, $V_h$, for type ab stars,
\be
V_h = V_{\rm F98} - 0.2 E(B-V) = 
V_H - 0.09 - 0.2 E(B-V)\qquad ({\rm RRab}).  \label{eqn:hipmag}
\ee
This change reduces the mean Hipparcos-based magnitudes
by only $\sim 0.01$ mag relative to the F98 calibration.

	We also follow F98 and remove three stars from
our previous sample (as defined in Paper II).  These are XZ Cet and AR Ser
which are anomalous Cepheids (not RR Lyrae stars) and BB Vir which shows 
evidence of having a horizontal-branch star companion.

	Figure \ref{fig:one}a shows the difference between the Hipparcos-based
magnitudes $V_h$ and the magnitudes of stars drawn by L96 from two catalogs:
B77 ({\it triangles}) and S91+S95 ({\it circles}).  
Both catalogs have been transformed according to L96.
The discrepancies are significant: 
$\langle V_{\rm L96} - V_h\rangle = -0.049\pm 0.009$ for B77 and
$\langle V_{\rm L96} - V_h\rangle = -0.100\pm 0.011$ for S91+S95, where
the errors are standard errors of the mean.
Figure \ref{fig:one}b shows the difference between the Hipparcos-based
magnitudes and the {\it original} B77 and S91+S95 
magnitudes for the same stars.  Here the agreement is excellent:
$\langle V_{\rm B77} - V_h\rangle = -0.006\pm 0.009$ and
$\langle V_{\rm S91,S95} - V_h\rangle = -0.014\pm 0.011$.   We conclude that
the original B77 photometry was on the modern scale and that
it was superior to the CD80 photometry.
Less surprisingly, the S91 and S95 CCD photometry is also on the modern scale.
We also show in Figure \ref{fig:one}b 16 type ab RR Lyrae stars with 
high-quality photometry ({\it squares}) from Jones et al.\ (1992), a subset
of which were used by F98 to test their Hipparcos-based magnitudes.  The
excellent agreement between Hipparcos and the solid points from 
Jones et al.\ (1992) and S91+S95 over the range $9.6\la V\la 12.3$ is strong
evidence that the Hipparcos magnitude scale is correct over the full range
of interest.  (Note that Jones et al.\ 1992 actually consider 17 type ab 
RR Lyrae stars, but we are unable to reconstruct the undereddened photometry
for one of these, SS Leo, from their sources, Liu \& Janes 1990 and 
Jones, Carney, \& Latham 1988.)

	Based on these results, we therefore revise the catalog of apparent
magnitudes as follows:  We adopt the Hipparcos-based magnitudes
of F98 (slightly adjusted for reddening as described 
by eq.\ \ref{eqn:hipmag})
whenever they are available.  For the stars without Hipparcos-based magnitudes,
we consider each of the six sources of photometry quoted by L96 separately.
We adopt the original (untransformed) photometry of B77 and S91+S95 since,
from Figure \ref{fig:one}b, these are in good agreement with Hipparcos.
L96 reported preliminary photoelectric photometry for eight stars, of which one
has a Hipparcos-based magnitude.  For the remaining seven stars, we
substitute the final values as reported by Layden (1997).  Layden (1997)
reports photoelectric photometry for one other star (V494 Sco) and
a photoelectric recalibration of another star (V413 Oph) with photographic 
photometry previously transformed from L94 (see below).  We adopt the Layden
(1997) values in both cases.  There are only 3 stars from Lub (1977) without
Hipparcos-based magnitudes (CP Aqr, AR Ser, and V494 Sco).  AR Ser is
an anomalous Cepheid (see above), and V494 Sco was remeasured by Layden (1997).
CP Aqr is a disk star, and so does not enter into our final results.  For
completeness, however, we note that we derived a transformation 
$V = V_{\rm L96}/0.985 - 0.116$ for the Lub (1977) stars using the 
Hipparcos-based magnitudes of four of them.  This leaves two sources:
CD80 and L94.

	Figure \ref{fig:three} shows $V_{\rm CD80} - V_h$ as a function
of $V_{\rm CD80}$ ({\it circles}).  For $V\la 10.5$,  CD80 is consistent with
Hipparcos, but for $11\la V\la 12$, CD80 is brighter, with a mean difference
of $-0.06\pm 0.01$ mag.  Thus, there is a clear trend with magnitude.
In principle, it is possible that this trend is due to systematic errors in
the Hipparcos-based photometry.  However, as we discussed above, there
is substantial evidence that Hipparcos is correct.  Since Hipparcos agrees
with B77, the same trend would appear in a comparison of CD80 and B77.
Indeed, it was this trend that L96 measured to derive their ``correction''
for B77 given by equation (\ref{eqn:bookmeyer}).  Since we established that
B77 is correct and that the trend is due to systematic errors in CD80, it
is appropriate to invert equation (\ref{eqn:bookmeyer}) to produce a 
correction for CD80,
\be
V =  {V_{\rm CD80} -0.146\over 0.983}.
\label{eqn:cd80}
\ee
This is shown as a dashed line in Figure {\ref{fig:three}.  Also shown
are the $V_{\rm CD80}$ magnitudes for the 18 CD80 stars without Hipparcos-based
magnitudes ({\it crosses}).  For 13 of these CD80 stars, there is also B77
photometry.  For these we plot the value of $V_{\rm CD80} - V_{\rm B77}$.
Clearly, these differences follow the pattern of the Hipparcos-CD80 
differences, confirming that B77 is on the Hipparcos system and CD80 is not.
Therefore, for the five CD80 stars without B77 or Hipparcos photometry
(shown with ordinate values of 0.2), we adopt the transformation given by 
equation (\ref{eqn:cd80}).  For the 13 CD80 stars with B77 photometry but
without Hipparcos photometry, we adopt the average of the B77 photometry and
the CD80 photometry as transformed by equation (\ref{eqn:cd80}).

	Figure \ref{fig:four} shows $V_{\rm L96} - V_h$ for 27 stars that 
were drawn from GVSC
by L94 and that have Hipparcos-based magnitudes.  These include 9 of the 39
such stars from L96 (the other 30 do not have Hipparcos-based magnitudes)
plus 18 additional stars that were not analyzed by L96.  
(A 19th star, IN Ara, has an Hipparcos proper motion but no Hipparcos
apparent magnitude because F98 found that the photometry
was of too low a quality to extract a reliable result.)\ \ 
For these 18 stars,
only L94 magnitudes were available.  L96 do not specify exactly how they
converted the L94 magnitudes, 
but by comparing the two catalogs, we find a very tight
relation $V_{\rm L96} = V_{\rm L94} + 0.0187(V_{\rm L94} - 10)$.  Stars
with photoelectric photometry are shown by crosses, and stars with
photographic photometry are shown by circles.  The six filled circles all 
derive from a single paper by Hoffmeister (1943) and are clearly
grossly in error.  These are all among the ``18 additional stars'' and so did 
not affect the L96 results, but did affect the results of Paper II.
Fortunately, none of the remaining 30 GCVS stars that lack
Hipparcos-based magnitudes
are drawn from this source, so it does not affect the present paper.  
Ignoring these six stars, the remaining 21 stars have
$\langle V_{\rm L96} - V_h\rangle =0.001\pm 0.027$, with a scatter of 
0.13 mag.  This subsample is therefore overall of substantially lower
quality than the rest of the sample.  Given the fact that it comprises only
$\sim 16\%$ of the full final sample (24 out of 147 stars), and given that the 
mean difference is 
close to zero, the quality would still seem to be acceptable.  
However, Figure
\ref{fig:four} has two additional, somewhat
disturbing, features.  First, the photographic measurements have substantially
larger scatter.  Second, the photographic measurements lie systematically
below zero, while the photoelectric measurements lie systematically above
zero.  Thus, we approach the subset drawn from GCVS through L94 (and without
Hipparcos-based magnitudes) cautiously.  The primary results that we report
incorporate these stars, but as a check, we also derive solutions by first
eliminating them.  

\section{Extinctions}

	Recent RR Lyrae statistical parallax studies have all relied 
primarily on the BH reddening map to account for extinction (L96; Papers I and
II;  F98).  We argue that this map should now be replaced
by the SFD map for three reasons.  

	First,  SFD is based on infrared emission
(as measured by the {\it COBE} and {\it IRAS} satellites) while BH is
based on 21 cm measurements of neutral hydrogen.  Infrared emission has
a direct physical relation to the dust, while HI is only indirectly related.
Moreover, the HI method can underestimate the dust in dense regions either
because the 21 cm line saturates or because HI is converted into molecular
hydrogen.  SFD give an instructive example of regions that look very
similar in HI, but have very different dust structures as measured by infrared
emission.

	Second, SFD covers more of the sky and does so in much greater detail.
Therefore it is bound to replace BH for most applications, including work
on the extragalactic distance scale.  To obtain the most reliable results,
one must measure the RR Lyrae absolute magnitude locally
and the apparent magnitudes
of RR Lyrae stars in external galaxies on the same system.

	Third, the SFD map and the BH map differ systematically in the
sense that SFD has 0.06 mag more extinction on average.  While there are
arguments in favor of both zero points, the arguments for the SFD
zero point appear more compelling to us.  In any event, even if the SFD
zero point were eventually proved wrong, such an error would, as noted in
the previous paragraph, cancel out in most distance-scale applications.
That is, the estimate of $M_V$ would be too bright, but the estimate
of the dereddened apparent magnitude
$V_0$ of RR Lyrae stars in external galaxies would be
too bright by the same amount, so the distance modulus $\mu = V_0 - M_V$ would
be unaffected.

	SFD establish the scale factor and zero point of their dust map
by finding the slope and intercept of the linear fit to the plot of
their dust measure against the observed $B-V$ color of elliptical galaxies.
Once this relation is fixed, every point on the sky is assigned an
extinction.  If the zero point of this relation were substantially in error,
then the point with the lowest extinction would most likely be either below 
zero or substantially above zero.  In fact, the lowest point 
is positive, but quite close
to zero, $A_V=0.015$.  If the BH zero point were correct, then the SFD map
should be adjusted downward by $0.06$ mag, and the best
estimate for the dust at this lowest point would be $A_V\sim -0.045$.
Thus, both the method for determining the SFD zero point and the actual result
appear very sound.

	In order to use the SFD map, we must initially restrict attention to
RR Lyrae stars that lie sufficiently far from the Galactic
plane that most of the dust
along the line of sight to extragalactic objects actually lies in front of
the star.  We choose a minimum distance from the plane of 
$z_{\rm min}=300\,$pc, and we assume that the dust has 
a scale height of
$h=130\,$pc.  Thus, for a star at height $z=300\,$pc, we assume that
$[1-\exp(-|z|/h)]\times 100\%=90\%$ 
of the dust lies in front of the star and hence that
the reddening $E(B-V)$ is 90\% of the value given by SFD.  We adopt
$R_V=A_V/E(B-V)=3.1$.  Of course, it is possible that along any given line
of sight all of the dust lies within 300 pc, and that along a few lines of
sight, a significant patch of dust lies beyond 300 pc.  However, from the
point of view of measuring the absolute magnitude of RR Lyrae stars, all that
is important is that this scale-height relation is correct on average.  

	Figure \ref{fig:five} shows L96 extinctions vs.\ SFD-based extinctions
for stars with $|z|>300\,$pc.
The SFD-based extinctions incorporate the correction for a dust scale height 
of $h=130\,$pc.  The L96 extinctions 
are based primarily on the BH system.  The diagonal line has
a slope of 1 and an intercept of $-0.05$ to account for the zero point 
differences between SFD and BH as determined from RR Lyrae stars.  
The solid triangles are ``halo-3'' stars
and the open triangles are disk stars.  The latter are not included in our
statistical parallax solution but are useful for studying systematics in
extinctions.  For $A_V\la 0.56$, the points are grouped
closely around the line, while at higher values they tend to fall below the
line.  We do not know the cause of this deviation, whether the SFD-base 
extinctions are too high or the BH-based extinctions are too low.  Plausible
arguments could be made either way.  In this paper we try to avoid all
possible sources of systematics, and we therefore eliminate the four stars
({\it circles})  with SFD extinctions $A_V>0.56$.

	The restrictions $|z|>z_{\rm max}$ and $A_V<0.56$ remove only 
about 10\% of the
sample.  Nevertheless, it would be nice to reincorporate these excluded stars, 
provided that there was a way to estimate their extinctions {\it on the
same scale} as the SFD map.  Alcock et al.\ (1998a,b) have measured the
correlation between log period $P_0$ and intrinsic $(V-K)_0$ color based on
SFD extinction estimates of a sample of 16 type ab RR Lyraes with well-measured
light curves.  They find,
\begin{equation}
(V-K)_0 = 1.052 \pm 0.013 + (1.059 \pm 0.147)(\log P_0 + 0.29).\label{eqn:vmkp}
\end{equation}
Unfortunately, of the 15 excluded stars in the ``halo-3'' statistical
parallax sample, we are able
to find mean $K$ magnitudes of sufficiently high quality for this purpose
for only one (TU Uma). 
We therefore use only stars with
$|z|>300\,$pc and $A_V<0.56$, which yields a sample of 147 stars of which
24 have photometry from GCVS.  (MS Ara is above 300 pc if its new Hipparcos
apparent magnitude is adopted and below 300 pc if its old GCVS magnitude is
used.  We include this star in all solutions to permit a fair comparison.)

\section{Results}

Table 1 shows the statistical parallax solutions for our final sample of
147 ``halo-3'' 
RR Lyrae stars (as defined in L96 and additionally specified in Paper II).
This sample is smaller than the 165 stars that we analyzed in Paper II 
primarily because of the removal of stars with poorly determined SFD
extinctions as discussed in \S\ 4, but also because of the removal of two
anomalous Cepheids and one star with a possible horizontal-branch companion
as discussed in \S\ 3.  We show four solutions: 1) without any changes
relative to the data used in Paper II, 2) with new apparent magnitudes
but with extinctions taken from Paper II, 3) with new (SFD) extinctions but
with apparent magnitudes taken from Paper II, and 4) with new extinctions
and new apparent magnitudes.  The numbers in parentheses below each row are 
the errors.  Note that while row (1) is not strictly
comparable to the results of Paper II because of the different number of
stars (147 vs.\ 165), the results are actually quite similar.  

	The first column is the distance-scaling factor, $\eta$, which
is normalized so that $\eta=1$ for the absolute-magnitude scale of L96,
\be
\eta\equiv 1 \Rightarrow M_V = 0.95 + 0.15{\rm [Fe/H]} \qquad {\rm (L96)}.
\label{eqn:L96def}
\ee 
Hence, the final result of incorporating the new apparent magnitudes and
new extinctions for 147 halo RR Lyrae stars is,
\be
M_V=0.77\pm 0.13,\quad {\rm at}\ \left< \rm [Fe/H]\right>=-1.60 \qquad 
({\rm pure}\ {\rm RR}\ {\rm Lyrae}).\label{eqn:pureRR2}
\ee

	Columns 2, 3, and 4, give the bulk motion of the sample in
$\kms$ relative to the local Galactic frame (with radial coordinate pointing
outward).  Here, the Sun is assumed to be moving relative to this frame at
 ${\bf v_{\odot}}=(-9,232,7) \;\rm{km} \, {\rm s}^{-1}$.  
The bulk motion is
$w_i=(6\pm 14,15\pm 13,1\pm 8)\,\kms$.  Columns 5, 6, and
7 give the square roots of the
diagonal terms of the velocity dispersion matrix, $C_{i j}^{1/2}$
in $\kms$.  The final row reports
$C_{i i}^{1/2}=
(\sigma_\pi,\sigma_\theta,\sigma_z) = (171\pm 11, 99\pm 8, 90\pm 7)\,\kms$.
We do not display the off-diagonal terms of $C_{i j}$ as we did in Papers I and
II because they are close to zero and uninteresting.

	All the results in Table 1 have been corrected for Malmquist bias,
scatter in the $M_V-$[Fe/H] relation, bias due to the anisotropic 
distribution of program stars on the sky, the effects of the non-Gaussian
RR Lyrae velocity distribution, and rotation from the heliocentric frame
to the local-Galactic frames of the stars, as discussed in detail in Paper I.

	As we discussed in \S\ 3, 24 of the 147 stars in this sample have
photometry from GCVS which is overall of lower quality (scatter $\sim 0.13$
mag) than that of the other 123 stars.  These stars should not be 
excluded on these grounds because this scatter is still small compared to
the intrinsic scatter of the method, $N^{1/2}\sigma_{M_V}= 1.57\,$mag,
where $N=147$ is the size of the sample and $\sigma_{M_V}=0.13\,$mag is the
error in equation (\ref{eqn:pureRR2}).  (See Appendix A of Paper II for an
additional discussion.)\ \ Nevertheless, for completeness we 
have determined the solution without these stars and find,
$M_V=0.77\pm 0.14$, almost exactly the same as equation (\ref{eqn:pureRR2}).

	In \S\ 4 we adopted $h=130\,$pc for the exponential scale height
of the dust.  We find that if we instead use $h=100\,$pc or $h=160\,$pc,
equation (\ref{eqn:pureRR2}) is changed by about 0.001 mag.  Even
if we were to use $h=300\,$pc, the change would be only 0.01 mag.

	In Paper II, we also considered a non-kinematically selected sample
of 827 stars with [Fe/H]$\leq -1.5$, including 103 RR Lyrae stars and
724 non-RR Lyrae stars from Beers \& Sommer-Larsen (1995).  
The available evidence suggests that RR Lyrae stars are similar in their
kinematics to other metal-poor stars (Ryan \& Lambert 1995; Chiba \& Yoshii 
1998), and that the kinematics of
metal-poor stars are not a strong function of metallicity 
(Beers \& Sommer-Larsen 1995).  Therefore the
non-RR Lyrae stars should define the same velocity ellipsoid as the RR Lyrae
stars, despite the fact that their mean metallicity is 
$\langle {\rm [Fe/H]}\rangle\sim -2.2$ compared to
$\langle {\rm [Fe/H]}\rangle\sim -1.8$ for the RR Lyrae sample.  Under this
assumption, the large number of non-RR Lyrae stars can independently
define the radial-velocity ellipsoid and so yield a check on the
radial-velocity ellipsoid (and hence the statistical-parallax solution) defined
by RR Lyrae stars alone.

	Here we repeat this procedure with a few changes.  First, we eliminate
all of the RR Lyrae stars that have questionable apparent magnitudes or
extinctions as discussed in \S\ 4.  Second, we
slightly change the selection criterion from [Fe/H]$\leq -1.5$ to
[Fe/H]$< -1.5$.  The reason for this is that L96 assigned to 7 RR Lyrae
stars for which there were no measured metallicities, an {\it assumed}
metallicity [Fe/H]$\equiv -1.5$, which was the mean for their entire sample.
One may show that for the pure RR Lyrae samples of L96, Paper I, Paper II,
and this paper,
this assumption introduces an utterly negligible random error of 0.002 mag.
Thus, the assumption is justified in order to maximize the size of the
usable sample.  However, incorporation of these stars could introduce a
bias for the {\it metallicity-selected} sample we are now considering.
By slightly changing the criterion to [Fe/H]$< -1.5$, we eliminate these
stars and very few others.  We then obtain a sample of 87 RR Lyrae stars
with $\langle {\rm [Fe/H]}\rangle = -1.81$
and 716 non-RR Lyrae stars with $\langle {\rm [Fe/H]}\rangle = -2.22$.  
Table 2 gives the results for this sample.
The description of the columns and rows is the same as for Table 1.
The results correspond to
\be
M_V=0.82\pm 0.13,\quad {\rm at}\ \left<\rm [Fe/H]\right>=-1.81 \qquad 
({\rm non-kinematic}),\label{eqn:nonkin}
\ee
$w_i=(4\pm 9,34\pm 9,2\pm 5)\,\kms$, and  
$C_{i i}^{1/2}=
 (\sigma_\pi,\sigma_\theta,\sigma_z) = (160\pm  7, 109\pm 8, 94\pm 5)\,\kms$.
Following the procedure
discussed in Paper II, we have added in quadrature a systematic error of
0.04 mag in going from Table 2 to equation (\ref{eqn:nonkin}) to take 
account of possibly different levels of thick-disk contamination in the RR 
Lyrae and
non-RR Lyrae solutions.  

	As in Paper II, we combine the two determinations (eqs.\
\ref{eqn:pureRR2} and \ref{eqn:nonkin})
taking account of the 0.44 correlation coefficient between them 
(see Appendix B to Paper II) and obtain,
\be
M_V=0.80\pm 0.11,\quad {\rm at}\ 
\left<\rm [Fe/H]\right>=-1.71 \qquad 
({\rm combined}).\label{eqn:combined2}
\ee
Note that equations (\ref{eqn:pureRR2}) and (\ref{eqn:combined2}) are in good
agreement with equations (\ref{eqn:pureRR}) and 
(\ref{eqn:combined}).

\section{Discussion}

	We take equation (\ref{eqn:pureRR2}) as the primary result of this
paper because the statistical-parallax solution for the pure RR Lyrae sample
requires essentially no additional assumptions.  By contrast, the 
non-kinematic solution of equation (\ref{eqn:nonkin}), and by implication
the combined solution of equation (\ref{eqn:combined2}), require the additional
assumption that metal-poor RR Lyrae stars have the same kinematics as
metal-poor non-RR Lyrae stars.  While this assumption is not absolutely
secure, there are a number of very strong arguments in its favor.  First,
as we discussed above and in Paper II, there is no evidence that RR Lyrae
and non-RR Lyrae stars can be distinguished kinematically.  Second, 
the available evidence suggests that
kinematics are independent of metallicity for [Fe/H] $<-1.5$.  Third, the 
pure RR Lyrae and non-kinematic solutions for $\eta$ agree within their
errors, even taking account of the 0.44 correlation coefficient between them
(see Tables 1 and 2).  Fourth, there is no statistically significant
difference between the {\it individual
velocity components} of the solution in Table 1 for 147 RR Lyrae stars and
the individual velocity components of the solution based {\it only} on the
716 metal-poor Beers \& Sommer-Larsen (1995) stars.  The latter differs 
somewhat from the solution in Table 2, and is given by
$w_i = (-2.2\pm 9.6, 38.3\pm 11.0,1.1\pm 5.5)\,\kms$ and
$C_{i i}^{1/2} = (160.0\pm 10.1 ,118.7\pm 13.1,92.6\pm 6.1)\,\kms$.  Taking
the difference between these six parameters and those in Table 1 and
dividing by the errors yields $(0.5, -1.4, -0.2, 0.8, -1.2, -0.3)$.
Of these six, only one ($w_\theta$) is possibly inconsistent with a normal
statistical fluctuation.  However, this is the {\it one} component that
we expect to be different because the pure RR Lyrae sample was selected
by removing stars with prograde orbits, while the non-kinematic sample was,
of course, selected without kinematic criteria.  Finally, if there were
any systematic difference between the RR Lyrae stars and the non-RR Lyrae
stars used in the non-kinematic sample, we would expect that it would be
in the sense of the non-RR Lyrae stars having more extreme kinematics
because they are on average more metal poor.  This would drive the
radial-velocity ellipsoid to higher dispersions and faster (relative to the
Sun) bulk motion, and hence would cause one to {\it overestimate} distances
(and luminosities) of the RR Lyrae stars when one attempted to match their
proper motions to these
high, non-RR Lyrae radial velocities.  That is, the only plausible bias
of this method is in the {\it same direction} as would be needed to resolve
the discrepancy between statistical parallax and other methods of determining
the absolute magnitude of RR Lyrae stars and {\it opposite in sign} from
the actual difference between the non-kinematic and pure RR Lyrae samples.
In brief, while equation (\ref{eqn:combined2}) does not sit on as firm
a foundation as equation (\ref{eqn:pureRR2}), it does argue very strongly
against the idea that equation (\ref{eqn:pureRR2}) is the
result of a large statistical fluctuation, particularly a fluctuation
in the direction of underestimating the RR Lyrae luminosity.

	As we discussed in the introduction, equation 
(\ref{eqn:pureRR2}) is in conflict at the $2\,\sigma$ level with the 
values determined from main-sequence fitting of clusters 
at the same metallicity of 
$M_V\sim 0.44\pm 0.08$ (Reid 1997) or $M_V\sim 0.49 \pm 0.04$ 
(Gratton et al.\ 1997; Gratton 1998).  
There are only four possible explanations for such a 
discrepancy: 1) a rare $(<1/20)$ statistical fluctuation, 2) a substantial
difference between cluster stars and field stars in the magnitude of the 
horizontal branch, 3) a systematic error in the main-sequence fitting distances
to clusters, or 4) a systematic error in the statistical parallax measurement.

	In this series of three papers, we have eliminated explanation (4).
Explanation (1) is of course always possible, but is unlikely.
Gratton (1998) has suggested explanation (2), that field and cluster 
horizontal branches might be different.  However, two lines of evidence
weigh against this possibility.  First, as Gratton (1998) notes, comparison of
the apparent magnitudes of RR Lyrae stars in LMC clusters with those of
neighboring field RR Lyrae stars (for which the reddening should be quite
similar) shows a mean offset of only $0.05\pm 0.02$.  However, this argument
strictly applies only to LMC RR Lyrae stars: there still could be a difference
between field and cluster RR Lyrae stars in the Galaxy which, unlike the
LMC, is a large spiral and probably has had quite a different
formation history.  In fact,  
Sweigert \& Catelan (1998) have produced models of two clusters 
with rising blue horizontal branches (NGC 6388 and NGC 6411) that have
RR Lyrae stars several tenths of a magnitude brighter than those of canonical
horizontal branch scenarios.  
The models invoke non-standard features, either high helium abundance,
high rotation velocity, or helium mixing at the tip of the giant branch, 
which cause the stars to have longer
periods at fixed temperature and metallicity.  Such long periods are actually
observed for the two known RR Lyrae stars in NGC 6388, enhancing the
plausibility of this explanation.  However, Catelan (1998) has shown that
the period-temperature diagrams for RR Lyrae stars in
five clusters that have been used for 
main-sequence fitting are actually quite similar in appearance to those of
field stars of similar metallicity.  Hence, while some cluster horizontal
branches may be brighter than those of the field, this does not appear to 
be the case for the clusters with main-sequence fitting distances.  

	We therefore consider that explanations (1), (2), and (4) are all 
rather
unlikely and that a systematic error in the main-sequence fitting distances is
the most plausible explanation for the discrepancy.  One possible cause of
a systematic error in the main-sequence fitting distances is that the
metallicities of the local subdwarfs might be on a different scale from
those of the clusters (determined from giants).  Specifically, if the
subdwarf metallicities were too low (or the giant metallicities too high)
then intrinsically brighter subdwarfs would be matched to the cluster main
sequences, leading to an overestimate of the cluster distance and of the 
luminosity of
its horizontal branch.  Recently, King et al.\ (1998) have found intriguing
evidence of a possible misalignment of this sort.  They measured the 
metallicities of M92 subgiants (not quite subdwarfs, but with higher gravities
than giants) and obtained metallicities up to half a dex lower than
those of M92 giants.  While there are a number of possible explanations for
this result, one is that the metallicities of giants are being systematically
overestimated or those of subdwarfs are being systematically underestimated.

\acknowledgements
We would like to thank A.\ Layden for clarifying a number of points
related to the data base he and his collaborators compiled, and for
sending us a copy of the GCVS reference list which was missing from
our library.
This work was supported in part by grant AST 94-20764 from the NSF.

\clearpage

\appendix
\renewcommand{\theequation}{\thesection\arabic{equation}}
\section{Metallicities}

	RR Lyrae luminosities are a function of metallicity and therefore
it is best if the mean metallicity of the calibrating stars be the
same as that of the stars whose distance one wants to measure.  Failing this,
one should at least know the difference between the two metallicities and
the slope of absolute magnitude with metallicity.

	Metallicity itself (in the sense of abundance of Fe relative to the 
solar value) is not an observable.  Rather it is a parameter derived from
observables (line strengths, colors, etc.) on the basis on atmosphere models.
In order for RR Lyrae stars to be effective distance indicators, it is
not necessary that their true metallicity be known.  Rather, as discussed
above, it is only necessary to know the {\it difference} in metallicities
between different samples.  In principle, this difference could be determined
directly from the observables.  In practice, the observables are usually
summarized as a single number: the ``metallicity''.  Thus, as long as two
metallicities are measured on the same ``system'' (and hence codify the
observables in the same way) the role of RR Lyrae stars as distance indicators
is not affected by absolute errors in the metallicity.  Metallicity 
therefore enters the statistical parallax calibration in a substantially
different way from the other observational inputs which is why we treat
it separately.  Of course, it is of some interest to know what the true 
metallicities of RR Lyrae stars are, and consequently what are their 
absolute magnitudes as a function of true metallicity.  However, this question
does not bear directly on the distance scale.

	The L94 metallicities (which we have adopted) are pinned
to the Zinn \& West (1984, ZW84) metallicity scale for globular clusters.  
(See, e.g., Fig.\ 1 of L94.)\ \ 
Hence, our statistical parallax calibration has the ZW84 scale embedded in it.
When this calibration is applied to other RR Lyrae stars with metallicities
on the ZW84 scale (such as Reid 1997), one can make a direct comparison
regardless of whether the ZW84 scale is actually correct.  However, there
are numerous other metallicity scales which can differ by several tenths
of a dex from ZW84.  For example,  Jurcsik \& Kov\'acs (1996) find that
the L94 (i.e., ZW84) scale is related to the Suntzeff, Kraft, \& Kinman 
(1994) scale by
${\rm [Fe/H]}_S = 0.957{\rm [Fe/H]}_L + 0.200$.  Comparing individual
cluster metallicities of Gratton et al.\ (1997) with ZW84, we find that
the former are about 0.1 dex more metal rich at [Fe/H]$_{ZW84}\sim -2$ 
and about 0.25 dex more metal rich at [Fe/H]$_{ZW84}\sim -1.5$.  Hence,
when we compared our absolute magnitude calibration at [Fe/H]$_{ZW84}\sim -1.6$
with Gratton et al.\ (1997) and Gratton (1998, specifically eq.\ 15), 
we actually used their absolute magnitude
at [Fe/H]$=-1.4$.  Note that if we had made the comparison ``naively'' at
the Gratton (1998) metallicity of [Fe/H]$=-1.6$, the discrepancy between the
two methods would have been more severe by 0.04 mag.  For this case the 
comparison is relatively straight
forward.  However, in general, metallicities are often quoted without
making clear what scale they are on, so that it may be difficult to
carry out rigorous comparisons in some cases.  In any event, one should
always proceed cautiously whenever making any such comparison.

	For the comparison with Gratton (1998), the situation is made simpler
by the fact that Gratton's (1998) slope of $S = d M_V/d{\rm [Fe/H]}$ is 
well constrained.  However, if we consider comparing the statistical parallax
result (at [Fe/H]$_{\rm SP}=-1.60$) with the value obtained by another 
method at a different metallicity, [Fe/H]$_*$, then there is an additional
uncertainty in the comparison, even if the two metallicities are put on the
same scale.  This uncertainty is $\Delta S\Delta$[Fe/H], where
$\Delta {\rm [Fe/H]} = {\rm [Fe/H]_{SP}} - {\rm [Fe/H]}_*$ and 
$\Delta S \sim 0.15$ is the difference of plausible slopes reported in the
literature, $S=0.15$ to $S=0.30$.

\clearpage

\begin{figure}
\caption[junk]{\label{fig:one}
Difference between the $V$ magnitude of RR Lyrae stars as determined
by F98 from Hipparcos data (slightly adjusted according
to eq.\ \ref{eqn:hipmag}) and the $V$ magnitude for the
same stars as determined by several other sources.  Shown are stars
with magnitudes reported by B77 ({\it triangles}), S91+S95 ({\it circles}), and
Jones et al.\ (1992) ({\it squares}).  The last are shown in panel (b) only.
In panel (a) we show the differences between the Hipparcos-based mags and
the mags of B77 and S91+S95 {\it as transformed by L96} according to
equations (\ref{eqn:bookmeyer}) and (\ref{eqn:schmidt}) respectively.  There
is a clear offset and trend with magnitude.  In panel (b) we show the
differences between the Hipparcos-based mags and the mags {\it as originally 
reported by the various authors}.  Panel (b) shows that the original
mags of all three samples are on the same system as Hipparcos.  Moreover,
the fact that the solid points (representing high-quality
ground-based photometry) fall close to the zero line over the interval
$9.6 \la V \la 12.3$ shows that F98's Hipparcos-based
photometry is quite precise over the entire range of interest.
}
\end{figure}

\begin{figure}
\caption[junk]{\label{fig:three}
Difference between the $V$ magnitude of RR Lyrae stars as determined
by F98 from Hipparcos data (slightly adjusted according
to eq.\ \ref{eqn:hipmag}) and the $V$ magnitude for the
same stars as determined by CD80 ({\it circles}).  Also shown are
18 stars with photometry from CD80 but without Hipparcos photometry 
({\it crosses}).  For the 13 such stars for which photometry is also 
available from B77, the difference $V_{\rm CD80}-V_{\rm B77}$ is shown.
The remaining 5 stars have an ordinate of 0.2.  There is a clear trend
with magnitude for the CD80 stars and this trend is similar for both
the stars with Hipparcos photometry and with B77 photometry.  The 
{\it dashed line}
(\ref{eqn:cd80}) is the expected deviation of the CD80 stars from the
true $V$ mag based on inverting equation (\ref{eqn:bookmeyer}) taken from L96,
that is inverting L96's choice of the best available photometry and
assuming it was B77 rather than CD80.
}
\end{figure}

\begin{figure}
\caption[junk]{\label{fig:four}
Difference between the $V$ magnitude of RR Lyrae stars as determined
by F98 from Hipparcos data (slightly adjusted according
to eq.\ \ref{eqn:hipmag}) and the $V$ magnitude for the
same stars as determined by L96 by transforming photometry from GCVS.
(For 18 of the stars only L94 photometry was available, but we have
transformed these to the L96 system.)\ \ Shown are stars with
photoelectric photometry ({\it crosses}) and photographic photometry
({\it circles}).  The six ({\it solid circles}) are all from a single
source (Hoffmeister 1943).  If these are removed, the scatter is 0.13 mag,
which is of lower quality than Hipparcos, B77, S91+S95, and CD80 (adjusted
according to eq.\ \ref{eqn:cd80}), but is still acceptable.  Of the
147 stars in the final sample, 24 have only GVCS photometry.
}
\end{figure}

\begin{figure}
\caption[junk]{\label{fig:five}
Extinctions of L96 (based primarily on BH) versus extinctions
based on SFD (including a correction for a dust scale height of
$h=130\,$pc). Shown
are ``halo-3'' stars ({\it solid triangles}) and disk stars ({\it open
triangles}).  The diagonal line has a slope of 1 and an intercept of
$-0.05$, to account for the zero point difference between the two maps.  
The points
systematically fall below the line for $A_V\ga 0.56$.
We therefore eliminate the four halo-3 stars with SFD extinctions $A_V>0.56$
{(\it circles}).
}
\end{figure}

\clearpage

\end{document}